\documentclass[aip,jcp,notitlepage,twocolumn,reprint,floatfix]{revtex4-2}

\usepackage{graphicx}
\usepackage{dcolumn}
\usepackage{bm}
\usepackage{hyperref}
\usepackage{amsmath}
\usepackage{mathtools}
\usepackage{booktabs}
\usepackage{multirow}
\usepackage{comment}
\usepackage{siunitx}
\usepackage{natbib}
\usepackage{xcolor}
\usepackage{tikz}
\let\oldtheequation\theequation
\makeatletter
\def\tagform@#1{\maketag@@@{\ignorespaces#1\unskip\@@italiccorr}}
\renewcommand{\theequation}{(\oldtheequation)}
\makeatother

\newcommand{\appref}[1]{\hyperref[#1]{Appendix~\ref{#1}}}
\newcommand{\lit}[1]{Ref.~\onlinecite{#1}}

\begin{document}

\title{A comparison between the one- and two-step spin-orbit coupling approaches based on the \emph{ab initio} Density Matrix Renormalization Group}

\author{Huanchen Zhai}
\email{hczhai@caltech.edu}
\affiliation{Division of Chemistry and Chemical Engineering, California Institute of Technology, Pasadena, CA 91125, USA}

\author{Garnet Kin-Lic Chan}
\email{gkc1000@gmail.com}
\affiliation{Division of Chemistry and Chemical Engineering, California Institute of Technology, Pasadena, CA 91125, USA}

\date{\today}

\begin{abstract}
The efficient and reliable treatment of both spin-orbit coupling (SOC) and electron correlation is essential for understanding f-element chemistry. We analyze two approaches to the problem, the one-step approach where both effects are treated simultaneously, and the two-step state interaction approach. We report an implementation of the \emph{ab initio} density matrix renormalization group (DMRG) with a one-step treatment of the SOC effect which can be compared to prior two-step treatments on an equal footing. Using a dysprosium octahedral complex and bridged dimer as benchmark systems, we identify characteristics of problems where the one-step approach is beneficial for obtaining the low-energy spectrum. 
\end{abstract}

\maketitle

\section{Introduction}

f-element molecules and materials possess unique magnetic and photochemical properties\cite{lucaccini2014beyond,buenzli2015design}
and have been proposed as components of atomic-scale quantum information processing.\cite{gaita2019molecular,aguila2014heterodimetallic} \emph{Ab initio} electronic structure computation can in principle provide a route to designing  the desired f-element chemistry,\cite{bohme2019link} however, unlike in lighter element compounds, treating both the strong spin-orbit coupling (SOC) and open-shell electron correlation is essential to establish the qualitative electronic structure.\cite{rinehart2011exploiting} Commonly, one uses the complete active space self-consistent field (CASSCF) method to treat  electron correlation in the f-orbital shell, then treats the SOC interaction in the basis of  spin-pure many-electron states. This is the so-called state-interaction spin-orbit (SISO) or 2-step approach.\cite{malmqvist1989casscf,malmqvist2002restricted} 

Although CASSCF-SISO has successfully treated many f-shell (and d-shell) systems,\cite{roos2004main,vancoillie2007calculation,yadav2016kitaev,ungur2017ab} the approach has some limitations.
First, in systems with more than a single f-shell atom, the CAS space rapidly becomes too large for exact CASSCF. To address this, approximate correlated electron solvers, such as density matrix renormalization group (DMRG),\cite{white1992density,white1993density,chan2002highly,chan2004algorithm,chan2011density,legeza2003controlling,barcza2011quantum,baiardi2020density,brabec2021massively,mitrushenkov2001quantum,chan2016matrix,sharma2012spin,olivares2015ab,wouters2014chemps2,wouters2014density,keller2015efficient,keller2016spin,zhai2021low} stochastic heat bath configuration interaction (SHCI),\cite{holmes2016heat,sharma2017semistochastic,holmes2017excited,smith2017cheap}, auxiliary-field quantum Monte Carlo (AFQMC)\cite{zhang2003quantum,al2006auxiliary,suewattana2007phaseless,purwanto2008eliminating,purwanto2009excited,motta2018ab}, and full configuration interaction quantum Monte Carlo (FCIQMC)\cite{booth2009fermion,cleland2010communications,booth2013towards,blunt2017density} have been explored. Analogous to CASSCF-SISO, such approximate active space solvers can be used in the 2-step state interaction approach, yielding SOC related properties. DMRG-SISO is an example of such a scheme.\cite{roemelt2015spin,knecht2016nonorthogonal,sayfutyarova2016state,sayfutyarova2018electron}

Second, there is the potential for the 2-step treatment of SOC to be inefficient when the effects of SOC are large.\cite{mussard2018one} Consequently, 1-step approaches to the problem, for example via extensions of  CASSCF,\cite{ganyushin2013fully} DMRG,\cite{hoyer2022relativistic} SHCI,\cite{mussard2018one} and AFQMC\cite{eskridge2022ab} solvers, directly compute eigenstates of the interacting Hamiltonian with SOC. Such 1-step approaches trade the well-optimized and simpler implementation of the spin-free correlation problem with the ability to only compute the eigenstates of interest, and without the need to solve the ancillary state-interaction problem. However, although there are theoretical benefits to the formulation, besides a very recent study based on iterative configuration interaction (iCI),\cite{zhang2022soici} we have not found a detailed and quantitative comparison of the performance of the 2-step and 1-step SOC approaches with the same CAS solver, making it difficult to fairly assess the  merits of each approach. 

In this work, we report a new implementation of \emph{ab initio} DMRG with SOC and electron correlation treated on an equal footing. While our DMRG implementation can handle various relativistic Hamiltonians following an approach similar to the one described in earlier relativistic DMRG studies by Knecht and coworkers,\cite{knecht2014communication,battaglia2018efficient} in this work we focus on a treatment via the Spin-Orbit Mean-Field (SOMF) Hamiltonian. We compare the performance of the new 1-step implementation with the existing 2-step DMRG-SISO approach. Based on a theoretical and numerical analysis of the two, we identify a problem where the 1-step SOC approach shows unambiguous advantages. The identified regime should be largely independent of the choice of solver, and thus helps clarify the role of 1-step and 2-step approaches in modeling f-element chemistry. 

\section{Theory}

\subsection{The Spin-Orbit Mean-Field Hamiltonian}

The treatment of relativistic effects in heavy elements can be carried out at different levels of theory\cite{dyall2007introduction}. The most direct way is to solve the four-component Dirac-Coulomb-Breit equation. However, this is expensive in complex molecules. Alternatively, one can represent relativistic effects as a correction to non-relativistic quantum chemistry. Namely, we can use a two-component Hamiltonian of the form\cite{dyall2007introduction,reiher2014relativistic}
\begin{equation}
    \hat{H} = \hat{H}^{\mathrm{SF}} + \hat{H}^{\mathrm{SO}}
\end{equation}
where \( \hat{H}^{\mathrm{SF}} \) and \( \hat{H}^{\mathrm{SO}} \) are the spin-free and spin-orbit coupling terms. Each piece contains relativistic contributions called the scalar relativistic correction and spin-dependent relativistic correction, respectively. 
As $\hat{H}^{\mathrm{SF}}$ has the same form as the non-relativistic Hamiltonian, the structurally new piece is the spin-dependent relativistic correction. 
The explicit form of the two terms varies with different approximations and implementations. In some special cases, different choices of the relativistic corrections can lead to quantitatively different results.\cite{mussard2018one}
In this work, we use the Breit-Pauli (BP) version of $\hat{H}^{\mathrm{SO}}$ within the SOMF approximation;\cite{neese2005efficient} alternative types of $\hat{H}^{\mathrm{SO}}$ starting from the X2C framework\cite{li2014spin,liu2018atomic,zhang2022atomic} could also be considered.
However, we will not discuss the merits of different approximate corrections, but instead only focus on the accuracy and efficiency of the implementation at the correlated electron level.

To formulate the electron correlation problem, we work in an all-electron basis of self-consistent field (SCF) molecular orbitals. For this purpose,  we do not use a spin-dependent relativistic correction at the SCF level. The molecular orbitals \( \{ \phi_i(\mathbf{x}) \} \) are then spin-independent real functions. In the molecular orbital basis, the spin-free Hamiltonian resembles its non-relativistic counterpart\cite{helgaker2014molecular}
\begin{equation}
    \hat{H}^{\mathrm{SF}} := \sum_{ij} t_{ij}^{\mathrm{SF}} \ \hat{E}_{ij}
    + \frac{1}{2} \sum_{ijkl} v_{ijkl}^{\mathrm{SF}}\
    \hat{E}_{ijkl}
\end{equation}
where the singlet excitation operators are given by
\begin{equation}
\begin{aligned}
    \hat{E}_{ij} =&\ \sum_{\sigma} a_{i\sigma}^\dagger a_{j\sigma} \\
    \hat{E}_{ijkl} =&\ \sum_{\sigma\sigma'}
    a_{i\sigma}^\dagger a_{k\sigma'}^\dagger a_{l\sigma'}a_{j\sigma}
\end{aligned}
\end{equation}
and
\begin{equation}
\begin{aligned}
    t_{ij}^{\mathrm{SF}} =&\ \int \mathrm{d}\mathbf{x} \
    \phi_{i}^*(\mathbf{x}) \left( -\frac{1}{2}\nabla^2 - \sum_a \frac{Z_a}{r_a} \right)
    \phi_{j}(\mathbf{x}) \\
    v_{ijkl}^{\mathrm{SF}} =&\
    \int \mathrm{d} \mathbf{x}_1 \mathrm{d} \mathbf{x}_2 \ \frac{\phi_{i}^*(\mathbf{x}_1)\phi_{k}^*(\mathbf{x}_2)
    \phi_{l}(\mathbf{x}_2)\phi_{j}(\mathbf{x}_1)}{r_{12}}
\end{aligned}
\end{equation}
are the ordinary real-number-valued one- and two-electron integrals, respectively, with the following permutation symmetries
\begin{equation}
\begin{aligned}
    t_{ij}^{\mathrm{SF}} = &\ t_{ji}^{\mathrm{SF}} \\
    v_{ijkl}^{\mathrm{SF}} = &\ v_{jikl}^{\mathrm{SF}} =
    v_{ijlk}^{\mathrm{SF}} = v_{klij}^{\mathrm{SF}} \\
    = &\ v_{jilk}^{\mathrm{SF}} = v_{lkij}^{\mathrm{SF}} =
    v_{lkji}^{\mathrm{SF}} = v_{klji}^{\mathrm{SF}}
\end{aligned}
\end{equation}

 To appropriately account for the spin-independent relativistic effects for heavy elements, for the applications in this work we compute the molecular orbitals using the spin-free X2C Hamiltonian, i.e. we use the spin-free X2C one-electron integrals \( t_{ij}^{\mathrm{SFX2C}} \) in place of \( t_{ij}^{\mathrm{SF}} \).
Since the total and projected spin are good quantum numbers for the spin-free Hamiltonian
\begin{equation}
    \big[ \hat{H}^{\mathrm{SF}}, \hat{S}^2 \big]
    = \big[ \hat{H}^{\mathrm{SF}}, \hat{S}_z \big] = 0,
\end{equation}
the SU(2) symmetry and block-diagonal structure of the Hamiltonian can be utilized to accelerate the computation of the non-relativistic many-body problem. Examples of the usage of these symmetries in DMRG are discussed in \lit{sharma2012spin} and \lit{wouters2014chemps2}.

For the BP-SOMF spin-dependent Hamiltonian term, the explicit definitions can be written as
\begin{equation}
    \hat{H}^{\mathrm{SO}} := \sum_{ij}
        \mathbf{t}_{ij}^{\mathrm{SOMF}} \cdot \hat{\mathbf{T}}_{ij}
\end{equation}
where the triplet excitation operators are
\begin{equation}
\begin{aligned}
    \hat{T}_{ij,x} =&\ \frac{1}{2} \big( a_{i\alpha}^\dagger
        a_{j\beta} + a_{i\beta}^\dagger a_{j\alpha}\big) \\
    \hat{T}_{ij,y} =&\ \frac{1}{2\mathrm{i}} \big( a_{i\alpha}^\dagger
        a_{j\beta} - a_{i\beta}^\dagger a_{j\alpha}\big) \\
    \hat{T}_{ij,z} =&\ \frac{1}{2} \big( a_{i\alpha}^\dagger
        a_{j\alpha} - a_{i\beta}^\dagger a_{j\beta}\big) \\
\end{aligned}
\end{equation}
and
\begin{equation}
    \mathbf{t}_{ij}^{\mathrm{SOMF}} = 
        \mathbf{t}_{ij}^{\mathrm{SO}}
    +\sum_{kl}D_{kl} \bigg(
      \mathbf{v}_{ijkl}^{\mathrm{SO}}
    - \frac{3}{2} \mathbf{v}_{ilkj}^{\mathrm{SO}}
    - \frac{3}{2} \mathbf{v}_{kjil}^{\mathrm{SO}}
    \bigg)
\end{equation}
are the SOMF effective one-electron integrals,
with \( [D_{ij}] \) the one-particle density matrix and
\begin{equation}
\begin{aligned}
    \mathbf{t}_{ij}^{\mathrm{SO}} =&\ \frac{\alpha^2}{2}
        \int \mathrm{d}\mathbf{x} \
    \phi_{i}^*(\mathbf{x}) \sum_a \frac{Z_a \hat{\mathbf{L}}_a}{r_a^3} 
    \phi_{j}(\mathbf{x}) \\
    \mathbf{v}_{ijkl}^{\mathrm{SO}} =&\ -\frac{\alpha^2}{2}
    \int \mathrm{d} \mathbf{x}_1 \mathrm{d} \mathbf{x}_2 \ \frac{\phi_{i}^*(\mathbf{x}_1)\phi_{k}^*(\mathbf{x}_2)
    \hat{\mathbf{L}}_{12}
    \phi_{l}(\mathbf{x}_2)\phi_{j}(\mathbf{x}_1)}{r_{12}^3}
\end{aligned}
\end{equation}
the spin-orbit one- and two-electron integrals where \( \alpha \) is the fine structure constant and \( \hat{\mathbf{L}} \) is the orbital angular momentum operator.

Neither the total spin nor projected spin is conserved in \( \hat{H}^{\mathrm{SO}} \). Therefore, the block-diagonal sparse structure (with respect to spin quantum numbers) of the non-relativistic Hamiltonian is lost in \( \hat{H} \). In addition, the \( \mathbf{t}_{ij}^{\mathrm{SOMF}} \) matrix elements are complex numbers. 

\subsection{The 1- and 2-Step Approaches} \label{sec:1and2}

In many systems involving third and fourth row transition metals, the effect of the spin-dependent term is relatively small, thus \( \hat{H}^{\mathrm{SO}} \) can be thought of as a perturbation to the spin-free Hamiltonian. The 2-step or state interaction approach is based on this idea. In the 2-step approach, we first solve the spin-free many-electron eigenvalue problem\cite{sayfutyarova2016state}
\begin{equation}
    \hat{H}^{\mathrm{SF}} | \Psi^{\mathrm{SF}}_{S,k} \rangle
    = E_{S,k}^{\mathrm{SF}} | \Psi^{\mathrm{SF}}_{S,k} \rangle
\end{equation}
to obtain a small set of spin-pure low-energy eigenstates \( | \Psi^{\mathrm{SF}}_{S,k} \rangle \), where the subscript \( S \) labels the total spin of the state. Then the relativistic Hamiltonian \( \hat{H} \) is constructed and diagonalized in the basis of these spin-pure states. Namely, we consider the effective Hamiltonian with matrix elements
\begin{equation}
    \hat{H}_{S_iS_j,ij}^{\mathrm{eff}} = E_{S_i,i}^{\mathrm{SF}}
        \delta_{ij} \delta_{S_iS_j} +
        \langle \Psi^{\mathrm{SF}}_{S_i,i} | \hat{H}^{\mathrm{SO}}
        | \Psi^{\mathrm{SF}}_{S_j,j} \rangle
\end{equation}
and the SOC corrected energy spectrum is obtained by diagonalizing  the small effective Hamiltonian matrix
\begin{equation}
    \hat{H}^{\mathrm{eff}} | \Psi_m^{[2]} \rangle = E_m^{[2]} | \Psi_m^{[2]} \rangle
\end{equation}
where the superscript \( [2] \) denotes a quantity obtained from a 2-step treatment.

In contrast, the 1-step approach solves the many-electron eigenvalue problem for $\hat{H}$ directly\cite{ganyushin2013fully}
\begin{equation}
    \hat{H} | \Psi_m^{[1]} \rangle = E_m^{[1]} | \Psi_m^{[1]} \rangle
\end{equation}

In the weak SOC regime, the 2-step approach has the following characteristics: 

(i) Accuracy. The accuracy can be systematically improved by including more low-energy spin-pure states. For some problems, symmetry analysis of the spin-pure states can be utilized to reduce the size of the effective problem.\cite{sayfutyarova2016state}

(ii) Efficiency. Given an existing performant non-relativistic code, the implementation can be optimized with little effort. The most time-consuming parts  are obtaining eigenstates of the spin-free Hamiltonian and evaluating the one-particle triplet transition density matrix (1TTDM) between spin-free states. Both computations can reuse highly optimized non-relativistic quantum chemistry subroutines. In addition, the non-relativistic problem has   SU(2) symmetry and spatial symmetry with only real-number-valued integrals. Finally, when the spin-free states are labelled by their total spin, the 1TTDM has a band sparse structure, since
\begin{equation}
    \langle \Psi^{\mathrm{SF}}_{S_i,i} | \hat{H}^{\mathrm{SO}}
    | \Psi^{\mathrm{SF}}_{S_j,j} \rangle = 0\quad (|S_i-S_j|>1)
\end{equation}
These features all help to make the computations efficient.

(iii) Dynamic correlation. Non-relativistic dynamic correlation can be approximately included by shifting the diagonal elements of the effective problem.\cite{ganyushin2013fully}

(iv) Interpretability. The 2-step approach yields the connection between the spin-mixed and spin-pure states as a byproduct. This can be useful for visualization and analysis.

However, we emphasize that the above features of the 2-step approach require the SOC to be small (relative to other electronic effects). 
When SOC is strong, some of the advantages of the 2-step approach are lost. In such a setting, we have the following considerations:

(i) Accuracy. When \( \hat{H}^{\mathrm{SO}} \) is large, the eigenstates of \( \hat{H}^{\mathrm{SF}} \) may poorly approximate the eigenstates of the full two-component  Hamiltonian.\cite{rinehart2011exploiting} Then, the 2-step approach will converge slowly with respect to the number of spin-pure eigenstates. Some studies have shown that one may need thousands of states per spin multiplicity to obtain reliable results for certain systems.\cite{ungur2017ab} Slow convergence is a particular problem in larger active spaces where the CAS calculation is expensive. 

(ii) Numerical conditioning. In an iterative eigenvalue solver such as the Davidson solver, it is relatively easy to find the lowest eigenstates because the energy gaps are relatively large. When targeting the interior eigenvalues however, the gaps are likely to almost vanish. A significantly larger number of iterations must then be used to converge those roots.\cite{dorando2007targeted} In contrast, the spectrum of the full relativistic Hamiltonian may have much larger gaps when the SOC effect is large, and typically we are only interested in a small number of SOC eigenstates. The
 numerical conditioning can result in the eigenvalue problem of the full Hamiltonian being easier than the non-relativistic one, and the gain in efficiency from the small number of solver iterations in the 1-step approach may offset the increased cost of removing symmetries present in the 2-step approach. 

(iii) Generalisability. The 2-step approach involves two separate eigenvalue problems (the spin-free eigenvalue problem and the  state-interaction eigenvalue problem). This complicates the generalization to further computations, such as to obtain response properties. In contrast, the 1-step approach retains the basic theoretical many-body structure of the spin-free many-body computation, with the exception of loss of symmetry. Thus it is easier to derive gradient and response expressions as analogs of those used in the spin-free theory. 

Considering the above arguments, we can propose a theoretical regime where the 1-step approach should be computationally superior to the 2-step approach. In particular, the 1-step formalism should possess advantages when (i) SOC is strong, (ii) the number of SOC eigenstates required is small, (iii) the size of the active space is large, and (iv) when we wish to obtain properties in addition to the energy. 

\subsection{The DMRG Implementation}

To understand the relative performance of the 1-step and 2-step approaches in a system with both strong SOC and electron correlation, we have implemented both approaches using the DMRG algorithm as the CAS solver. As the treatment of dynamic correlation effects (such as through NEVPT2\cite{angeli2001introduction,angeli2002n}) in the 1- and 2-step calculations is not strictly comparable (see \autoref{sec:1and2}), we do not consider dynamic correlation outside the CAS space in this work.

The detailed description of the DMRG-SOSI (2-step) approach can be found in \lit{sayfutyarova2016state}. The main part of a DMRG-SOSI implementation is the spin-adapted \emph{ab initio} DMRG algorithm without the use of singlet embedding~\cite{sharma2012spin,li2017spin} (since it is not clear how one can use singlet embedding to compute the 1TTDM between states with different total spins). We re-implemented the DMRG-SOSI approach in the DMRG code \textsc{Block2}\cite{block2} using atomic integrals and SCF solutions computed using \textsc{PySCF}.\cite{sun2018pyscf,sun2020recent} 

The 1-step approach requires extending the DMRG code to support complex number arithmetic. In addition, for the 1-step approach, it is advantageous to work in a general spin orbital formalism. We use a matrix product operator (MPO) in the general spin orbital basis in our implementation. Based on this, all normal/complementary operator indices are spin orbital indices. The DMRG formulae for a general spin orbital implementation can be found in some other publications.\cite{chan2002highly,chan2016matrix,ren2020general}

The simplest implementation stores all MPO and matrix product state (MPS) data as complex numbers. However, since the two-body part of the MPO and the rotation matrices in the MPS are actually real-number-valued, there will be many exact zeroes in the imaginary part of their representation. Using the floating point number compression method introduced in \lit{zhai2021low}, these zeroes incur only a negligible amount of disk storage. Nevertheless, the memory and computational costs do not take these zeroes into account. Ideally, one would use the sub-Hamiltonian approach\cite{chan2016matrix,zhai2021low} to represent the spin-free part and (pure imaginary part of) the spin-orbit piece of the Hamiltonian as two independent sub-MPOs. Then we could represent the first sub-MPO using only real numbers, reducing memory and computational costs while keeping the accuracy unchanged, since we only require complex-valued computations for the second sub-MPO which does not involve two-electron integrals (when SOMF is used).
In practice, however, we found the efficiency of the hybrid scheme to depend greatly on some implementation details. Thus in this work, we report results from the 1-step approach using  the single complex MPO, with the MPO automatically constructed using the bipartite matching scheme described in \lit{ren2020general}.

\section{Numerical Example}

\subsection{d-element Systems}

To benchmark the accuracy of our new SOC-DMRG implementation, we first consider some d-element atoms. We computed the zero-field splitting (ZFS) between the \( {}^2\mathrm{S}_{1/2} \) and \( {}^2\mathrm{D}_{5/2} \) and \( {}^2\mathrm{D}_{3/2} \) states for the Cu and Au atoms,\cite{sayfutyarova2016state} using the ANO-RCC basis\cite{roos2005new} contracted to 6s5p3d2f for Cu and 8s7p4d2f for Au, respectively. We performed the CASSCF calculation with an 11 electron, 11 orbital active space and including 6 doublets in the state-averaged treatment and the 2-step DMRG. We list the results in \autoref{tab:cuau}. The computed splittings   agree well with previous theoretical studies and their experimental values.

\begin{table}[!htbp]
    \centering
    \caption{ZFS for the Cu and Au atoms computed using DMRG-CASSCF (in eV).}
    \label{tab:cuau}
    \begin{tabular}{
        p{2.0cm}
        >{\centering}p{1.4cm}
        >{\centering}p{1.4cm}
        >{\centering}p{1.4cm}
        c}
        \hline\hline
        term & 1-step SOC & 2-step SOC & 2-step SOC
        \cite{sayfutyarova2016state} &
        experiment\cite{sansonetti2005handbook}  \\
    \hline
    \\
    \multicolumn{5}{c}{Cu atom with CASSCF(11e, 11o) orbtials} \\
    \\
    \( {}^2\mathrm{D}_{5/2} \) & 1.568 & 1.568 & 1.57 & 1.39 \\
    \( {}^2\mathrm{D}_{3/2} \) & 1.829 & 1.828 & 1.83 & 1.64 \\
    \( {}^2\mathrm{D}_{5/2}-{}^2\mathrm{D}_{3/2} \)
                               & 0.261 & 0.260 & 0.26 & 0.25 \\
    \\
    \multicolumn{5}{c}{Au atom with CASSCF(11e, 11o) orbtials} \\
    \\
    \( {}^2\mathrm{D}_{5/2} \) & 1.679 & 1.652 & 1.68 & 1.14 \\
    \( {}^2\mathrm{D}_{3/2} \) & 3.416 & 3.365 & 3.39 & 2.66 \\
    \( {}^2\mathrm{D}_{5/2}-{}^2\mathrm{D}_{3/2} \)
                               & 1.737 & 1.714 & 1.71 & 1.52 \\
    \\
    \hline\hline
    \end{tabular}
\end{table}

\subsection{f-element Systems}

\begin{figure}[!htbp]
  \includegraphics[width=\columnwidth]{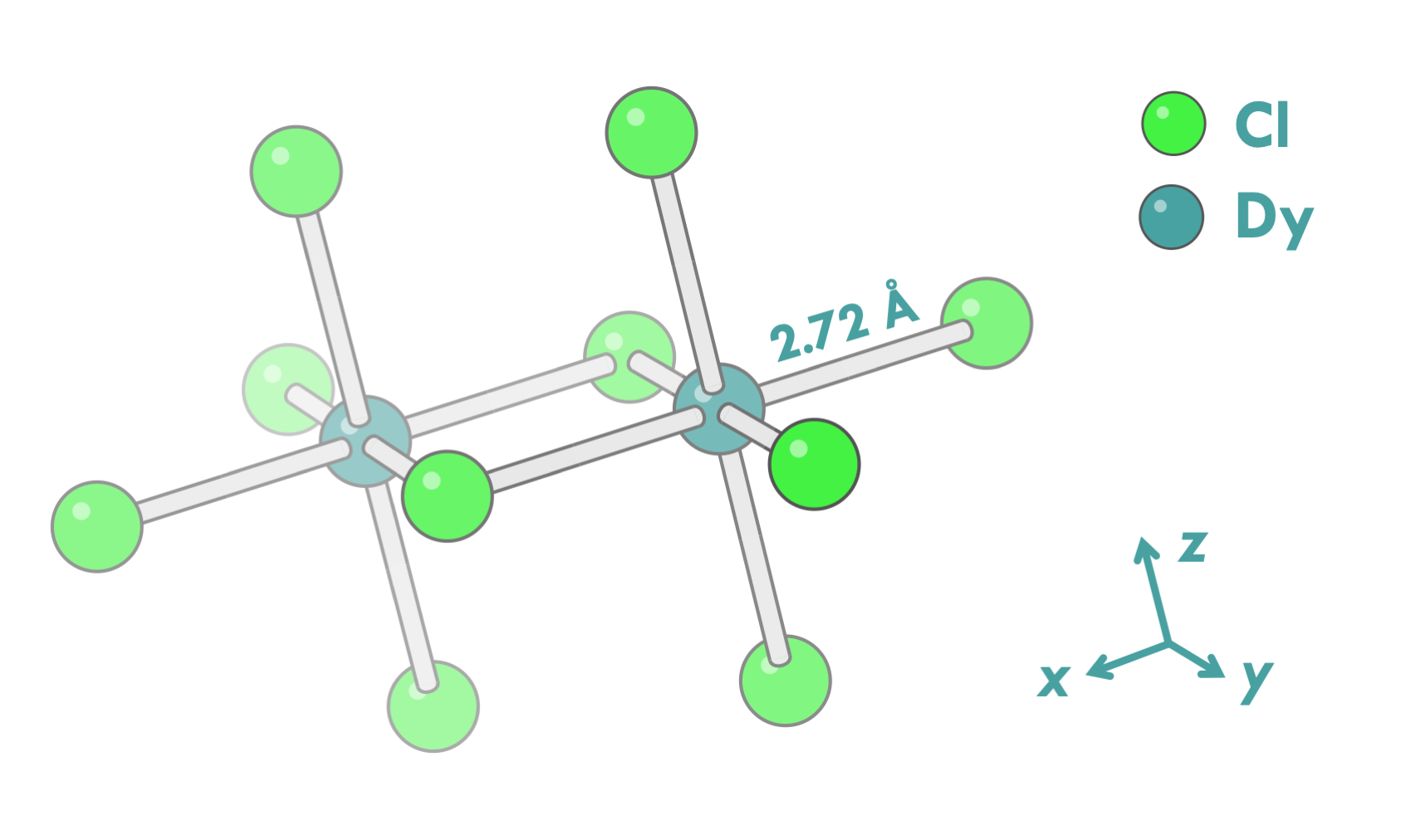}
  \caption{The geometry of the dysprosium dimer complex \( \mathrm{[Dy_2Cl_{10}]^{4-}} \). }
  \label{fig:dimer}
\end{figure}

To further investigate the performance of the 1- and 2-step DMRG-SOC implementations, we consider an artificial f-element molecule, the  edge-sharing (bridged) dysprosium dimer complex \( \mathrm{[Dy_2Cl_{10}]^{4-}} \). Each Dy atom is in an octahedral crystal field, shown in \autoref{fig:dimer}. The corresponding monomer \( \mathrm{[DyCl_6]^{3-}} \) has been studied in \lit{aravena2016periodic} where the standard CASSCF / NEVPT2\cite{angeli2001introduction,angeli2002n} 2-step approach was used. In this work, we set the \( \mathrm{Dy^{III}-Cl^-} \) bond length to 2.72 \AA\  according to Table 3 in \lit{aravena2016periodic}, which is the optimized bond length found in the octahedral \( \mathrm{Dy^{III}Cl_6Na_6} \) model.

For the mean-field calculations, we first performed an unrestricted Kohn-Sham (UKS) calculation with the BP86 functional.\cite{perdew1986density,becke1988density} The Dy and Cl atoms were described by  the ANO-RCC basis set\cite{roos2005new}, contracted to 7s6p4d2f for Dy and 4s3p for Cl, respectively. The mean-field model then has 306 electrons in 248 spatial orbitals in the dimer case. To generate the orbital space for the DMRG calculations, we performed UKS with the high spin \( S_z = 5/2 \) (monomer) or \( S_z = 5 \) (dimer) state. The obtained orbitals were split-localized using the Pipek-Mezey algorithm.\cite{pipek1989fast} The selected CAS configuration interaction (CASCI) space consisted of 30 electrons in 20 spatial orbitals (i.e. 40 spin orbitals) containing the 14 Dy 4f and 6 bridging Cl p orbitals for the dimer, or 9 electrons in 7 Dy 4f spatial orbitals for the monomer. For the monomer case, we optimized the orbitals using state-averaged CASSCF over 42 doublets, 42 quartets, and 21 sextets, while for the dimer case the orbitals were not optimized.

We determined the ordering of the orbitals used in the 1- and 2-step DMRG using the Fiedler\cite{olivares2015ab} scheme. We used the state-averaged DMRG algorithm to obtain the ground and excited states. Note that the state-averaged algorithm over a given number of states $N_\text{av}$ does not necessarily converge to the lowest $N_\text{av}$ states (for example, depending on the initial guesses, some excited states might be missed) but will converge to a set of eigenstates at large DMRG bond dimension.

For the dimer case where a large number of states were included in the state-averaged treatment, we further refined the eigenstates using the level-shifted Hamiltonian

\begin{equation}
    \hat{H}'_k = \hat{H} + \sum_{i=1}^{k-1} w_i |\Psi_i\rangle\langle \Psi_i|
\end{equation}
where \( |\Psi_i\rangle \) are known (refined) states with energies below the targeted excited state \( |\Psi_k\rangle \), and \( w_i \) are the weights (energy level shifts). In this work, we set \( w_i = 0.5 \ \mathrm{Hartree} \). The additional energy gain from refinement was quite small for the 2-step approach; in the spin-pure case, the improvement from refining the state energies was mostly less than \( 30 \ \mathrm{cm}^{-1} \), with a few cases near \( 50 \ \mathrm{cm}^{-1} \), when averaging over 36 multiplets in each multiplicity. Consequently, to  reduce the total wall time for the 2-step approach, the 1TTDM was computed using the state-averaged states, rather than state-specific refined states. For the 1-step approach, we report the energies after this refinement.

The UKS calculations with orbital localization were performed using \textsc{PySCF}\cite{sun2018pyscf,sun2020recent} with some helper functions from \textsc{libDMET}.\cite{cui2022systematic} The SOMF integrals were obtained from \textsc{PySCF}. All DMRG calculations were performed in \textsc{Block2}.\cite{zhai2021low} The calculations were executed on nodes with 28-core Intel Cascade Lake CPUs (2.2 GHz), made available via the Caltech high-performance computing facility. Each node has 56 CPU cores and 384 GB of memory. For the dimer calculations, each DMRG job used 1 or 2 nodes. For the 2-step approach the main parallelism is over computations of states with different spin multiplicities. We set the MPS bond dimension \( M = 2000 \) in all DMRG calculations, and the final discarded weight is below \( 1\times 10^{-5} \) for both the 1- and 2-step calculations. 

\subsubsection{Low-Energy Spectra}

\begin{figure}[!htbp]
  \includegraphics[width=\columnwidth]{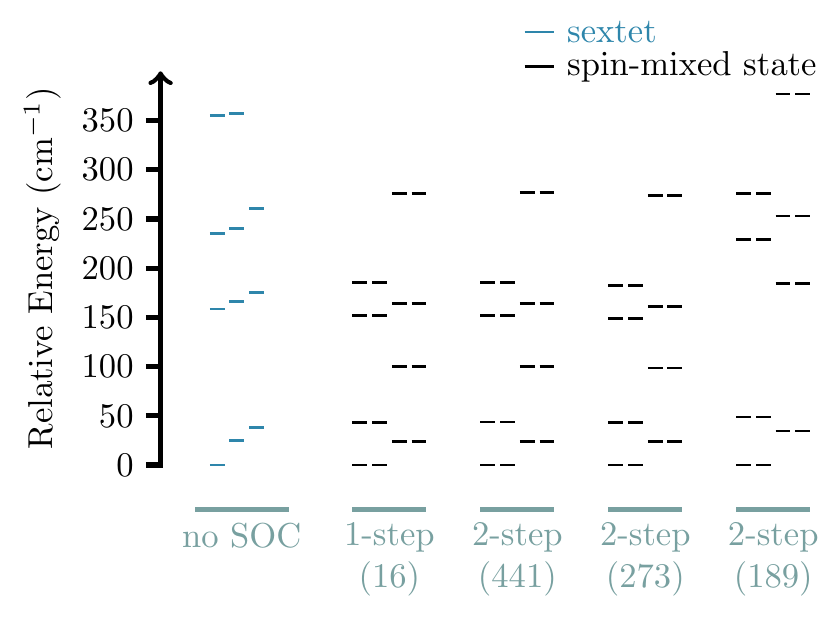}
  \caption{Low-energy spectrum of \( \mathrm{[DyCl_{6}]^{3-}} \) with CAS(9e, 7o) obtained from DMRG without SOC and 1- and 2-step DMRG-SOC approaches. Spin multiplets and spin-mixed states are shown as colored and black bars, respectively. Arbitrary horizontal shifts are used to separate the near-degenerate states. The numbers in  parentheses indicate the total number of states used in the state-averaged DMRG in the 1-step approach or the first step of the 2-step approach.}
  \label{fig:spec-mono}
\end{figure}

\begin{figure}[!htbp]
  \includegraphics[width=\columnwidth]{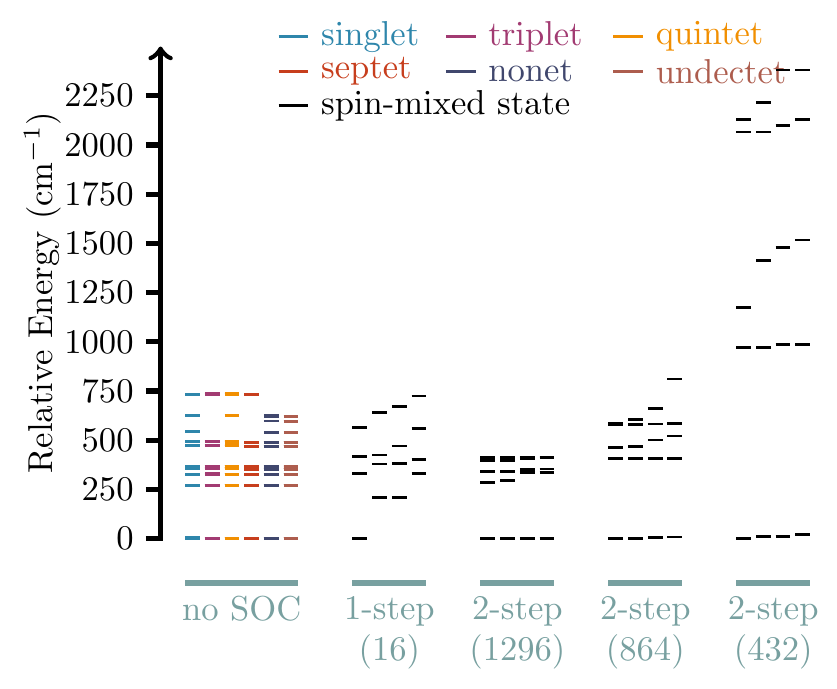}
  \caption{Low-energy spectrum of \( \mathrm{[Dy_2Cl_{10}]^{4-}} \) with CAS(30e, 20o) obtained from DMRG without SOC and 1- and 2-step DMRG-SOC approaches. Notation is the same as in the previous figure.}
  \label{fig:spec}
\end{figure}

We plot the low-energy spectra of the dysprosium monomer and dimer complexes studied in this work in \autoref{fig:spec-mono} and \autoref{fig:spec}, respectively. For the no-SOC case, each single level represents an entire spin multiplet. The lowest 11 and 216 multiplets, corresponding to 66 and 1296 eigenstates, are shown for the monomer and dimer respectively. For the 1- and 2-step approaches, the lowest 16 (or 20) spin-mixed eigenstates with SOC corrected energies are shown for the monomer and dimer. As the 1-step approach energies are converged with respect to the DMRG bond dimension, they can be considered to provide reference SOC energies for the eigenstates (although as discussed above, it is not guaranteed that the set of eigenstates are the lowest set). For ease of comparison, the ground-state energies are shifted to a common zero. 

\definecolor{sing}{RGB}{46, 134, 171}
\definecolor{trip}{RGB}{162, 59, 114}
\definecolor{quin}{RGB}{241, 143, 1}
\definecolor{sept}{RGB}{199, 62, 29}
\definecolor{nont}{RGB}{64, 71, 109}
\definecolor{unde}{RGB}{173, 93, 78}
\definecolor{sext}{RGB}{46, 134, 171}

In the monomer, the large SOC is reflected in the very different number of states at low energy in the no-SOC data versus the SOC spectra. In the case of no-SOC, there are only 6 states below \(200\ \mathrm{cm}^{-1}\) (all from sextets), while with SOC there are 14 states in this energy window. The ground-state shift due to SOC is \( -6375\  \mathrm{cm}^{-1} \) (not shown in the figure).
To obtain qualitative agreement between the 2-step and 1-step spectra for the lowest 16 states, we find that we need to use 42 doublets, 42 quartets, and 21 sextets (273 spin-free states in total) in the 2-step state-interaction problem. When we use 84 doublets, 84 quartets, and 21 sextets (441 spin-free states in total), we find excellent agreement between the 2-step and 1-step spectra. Note that for this CAS(9e, 7o) problem, the full spin-free spectrum contains 490 doublets, 224 quartets, and 21 sextets.

In the dimer, the no-SOC spectra for different spin multiplicities are very similar. This indicates that the Heisenberg \( J \) coupling between the two metal centers is very weak. Because of this, obtaining a precise value for $J$ requires a more detailed treatment of dynamical correlation and its balance between different states, which is outside the scope of this work. Above \( 200\ \mathrm{cm}^{-1} \), the density of states is very high (not fully shown in the figure). Each multiplicity has the same number of states in the state-average, and because of the small $J$ coupling, we would expect all the eigenstates to have similar energies. However, we observe that the state-averaged calculations for different multiplicities do not yield all similar energy levels (and thus miss some of the eigenstates). 
The high density of excited states in the no-SOC spectra not only makes the spin-free calculation harder to converge but also introduces difficulties in selecting spin-free states for the state interaction treatment, since many states can contribute similarly in the SOC treatment.

Similarly to the monomer, the SOC-corrected spectra are quite different from the no-SOC spectra, only more so due to the high density of states in the no-SOC spectrum. In fact, the state-averaged 1-step calculations find only 11 spin-mixed states within the energy range of \( 500 \  \mathrm{cm}^{-1} \) above the ground-state, as compared to 1152 spin-free states. The SOC correction for the absolute energy of the ground state is \( -12723 \  \mathrm{cm}^{-1} \) (not shown in the figure). 
 
To obtain  a similar result to the reference 1-step spectrum using the 2-step approach, we mainly considered 3 different settings, with 12, 24, and 36 multiplets per multiplicity (possible multiplicities were 1,3,5,7,9, and 11) used in the state interaction treatment, respectively. These choices generated in total 432, 864, and 1296 spin-free states. \autoref{fig:spec} illustrates that due to the insufficient number of spin-free states used in the 2-step treatment, the difference between the 1- and 2-step spectra is very large.

\subsubsection{Convergence}

\begin{figure}[!htbp]
  \includegraphics[width=\columnwidth]{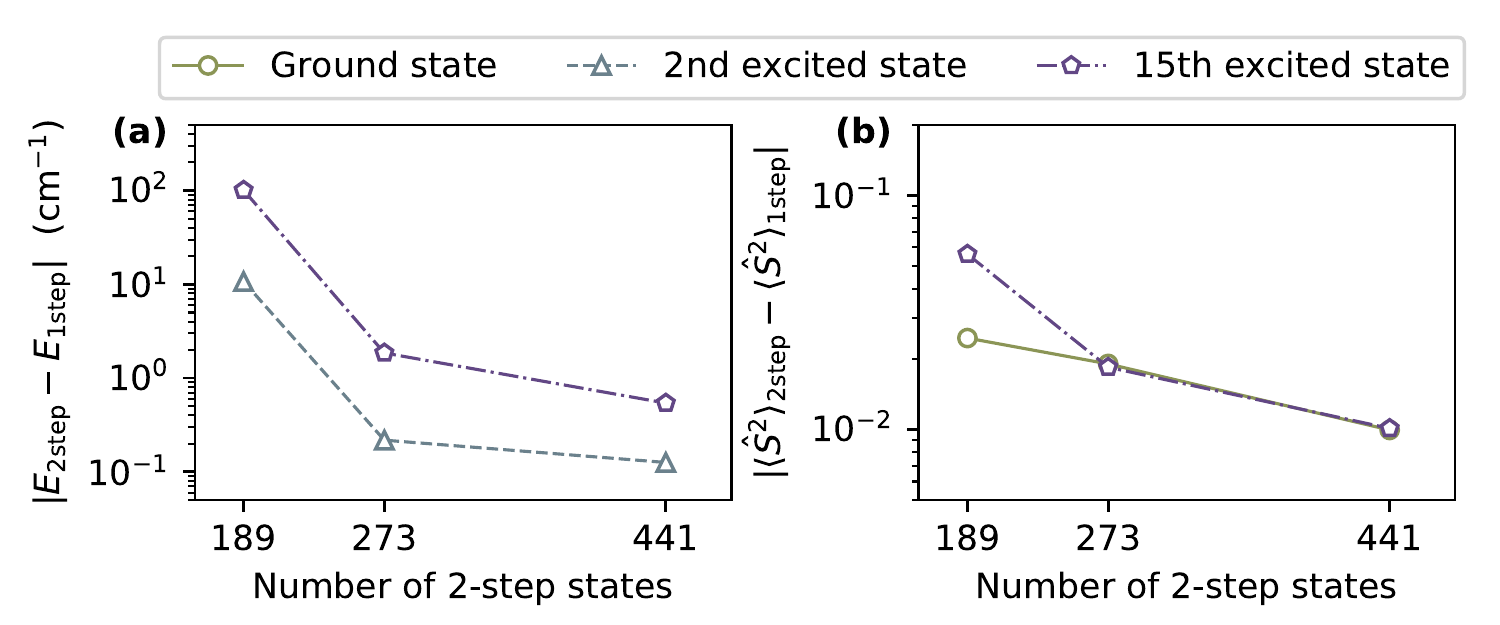}
  \caption{The difference between representative low-energy states computed from 1-step and 2-step approaches, in (a) excitation energy and (b) \( \langle \hat{S}^2 \rangle \) for the monomer.}
  \label{fig:error}
\end{figure}

\autoref{fig:error} shows the convergence of the excitation energies and squared total spin \( \langle \hat{S}^2 \rangle \) from 2-step to 1-step for the monomer. We see that the convergence with respect to the number of states included in the 2-step approach is faster for the energy than  for  \( \langle \hat{S}^2 \rangle \). From the 1-step calculation, we see that the monomer ground state \( \langle \hat{S}^2 \rangle \) is 8.48, which is very close to that of the pure sextet (\( \frac{5}{2}(\frac{5}{2} +1) = 8.75\)).
In the dimer case, there is no systematic convergence in the 2-step approach as the number of states is increased. This is seen in Table~\ref{tab:dimerspins}, which shows the \( \langle \hat{S}^2 \rangle \) values for the lowest 5 states as a function of the number of averaged states in the 2-step approach. Additional information on the energies and \( \langle \hat{S}^2 \rangle \) values for the lowest states can be found in the Supporting Information.

\begin{table}[!htbp]
    \centering
    \caption{\( \langle \hat{S}^2 \rangle \) computed from 1-step and 2-step approaches for the lowest five states of the dimer. Note that the results from 2-step calculations with more than 1296 spin-pure states may not be fully converged.}
    \label{tab:dimerspins}
    \begin{tabular}{p{2.5cm}
        S[table-format=3.2]
        S[table-format=3.2]
        S[table-format=3.2]
        S[table-format=3.2]
        S[table-format=3.2]}
    \hline\hline
    SOC scheme & \multicolumn{5}{c}{\( \langle \hat{S}^2 \rangle \)} \\
    \hline
    \\
    1-step (16) & 28.9 & 27.3 & 26.6 & 25.1 & 25.2 \\
    \\
    2-step (432) & 4.98 & 30.0 & 30.0 &  5.01 & 29.8 \\
    2-step (864) & 29.9 & 29.9 &  5.07 &  5.06 & 18.9 \\
    2-step (1296)& 29.9 & 29.9 &  5.09 &  5.09 &  7.50 \\
    2-step (1728)& 29.9 & 29.9 & 5.05 & 5.09 & 7.74\\
    2-step (2592)& 30.0 & 30.0 & 29.6 & 29.6 & 25.9 \\
    \\
    \hline\hline
    \end{tabular}
\end{table}

\subsubsection{Efficiency}

We list timings for the dimer DMRG calculations performed in this work in \autoref{tab:timing}. For the 2-step approach, we performed DMRG calculations for each spin multiplicity separately, thus the listed CPU hours are the sums of the CPU hours from all multiplicities. Note that the timings can depend on details of the DMRG parameters and implementation. Also the meaning of the MPS bond dimension \( M \) for the general spin MPS and spin-adapted MPS is different, so an absolutely fair comparison is quite difficult. Nevertheless, we see that the 1-step state-averaged DMRG calculations generally incur significant overhead over the 2-step state-averaged calculations. This is mainly because we have to use complex-valued MPS and a general spin MPO for the 1-step Hamiltonian.
Nonetheless, when computing 16 states in the 1-step approach, the cost was of the same order of magnitude as that of the spin-free part of the 2-step approach computing hundreds of states. However, the second part of the 2-step approach requires computing the 1TTDM.
The computational cost to obtain the 1TTDM increases rapidly with the number of interacting states. Therefore, we find that the total CPU cost required for a reliable 2-step calculation is in fact significantly greater than that required for a 1-step calculation.

\begin{table}[!htbp]
    \centering
    \caption{Measured and estimated timings for the 1- and 2-step approaches for the dimer.}
    \label{tab:timing}
    \begin{tabular}{
        >{\centering\arraybackslash}p{2.7cm}|
        >{\centering\arraybackslash}p{1.3cm}
        >{\centering\arraybackslash}p{1.1cm}
        >{\centering\arraybackslash}p{1.1cm}
        >{\centering\arraybackslash}p{1.1cm}
        }
        \hline\hline
        total CPU hours &
    1-step (16)& 2-step (1296) & 2-step (864) & 2-step (432) \\ 
    \hline
    DMRG & 6889 & 7487 & 5086 & 2958 \\
    1TTDM & 0 & 2039 & 907 & 206 \\
    \hline\hline
    \end{tabular}
\end{table}

\section{Conclusions}

In this work we carefully analyzed the relative strengths and shortcomings of 1-step and 2-step approaches to treating spin-orbit coupling, using the context of a new DMRG implementation as an example. In numerical tests on a dysprosium dimer complex, we showed that the 1-step approach is preferred when computing the low-energy spectrum due to the strong spin-orbit coupling and high density of states. In particular, the 2-step approach converges very slowly with   the number of included spin-free states. 
For less symmetric systems, the density of states may be lower than the example considered in this work, and this may alleviate some of the problems of the 2-step approach even when the SOC is strong.

In problems of electronic structure with a large amount of degeneracy in the spin-free spectrum, strong SOC can split the degeneracy. In those cases, treating the SOC and non-dynamic correlation simultaneously may be easier since the spin-mixed states will be well separated from each other.
The remaining problem then becomes how to efficiently compute with the more complicated Hamiltonian. As shown in this work in the case of the DMRG implementation, there remains significant overhead when working with two component Hamiltonians. This may be addressed in future work. 

\begin{acknowledgments}
This work was supported by the US Department of Energy, Office of Science, via award DE-SC0019390. HZ thanks Zhi-Hao Cui for useful discussions on mean-field calculation and high performance computing strategies, and Xubo Wang for discussions on relativistic Hamiltonians and spin-orbit effective core potentials. The computations presented in this work were conducted at the Resnick High Performance Computing Center, a facility supported by the Resnick Sustainability Institute at the California Institute of Technology.
\end{acknowledgments}

\section*{Author Declarations}
The authors have no conflicts to disclose.

\section*{Data Availability}
The data presented in this work can be reproduced using the open-source \textsc{PySCF} 2.0.1,\cite{sun2018pyscf,sun2020recent} \textsc{Block2} 0.5.1,\cite{block2} and \textsc{libDMET} 0.4\cite{libdmetsolid} codes. The reference input and output files can be found in the GitHub repo \url{https://github.com/hczhai/dmrg-soc-data}.

\bibliography{main}

\end{document}